\documentclass[%
 reprint,
 amsmath,amssymb,
 aps,
 prl,
floatfix,
]{revtex4-2}

\usepackage{graphicx}
\usepackage{dcolumn}
\usepackage{bm}
\usepackage{hyperref}

\begin{document}

\title{Mobility Census for monitoring rapid urban development}

\author{Gezhi Xiu$^{1,3,*, \dag}$, Jianying Wang$^{2, \dag}$, Thilo Gross$^{4,5,6}$, Mei-Po Kwan$^2$, Xia Peng$^7$, Yu Liu$^1$}
\affiliation{$^1$Institute of Remote Sensing and GIS, Peking University,\\
$^2$Institute of Space and Earth Information Science, The Chinese University of Hong Kong (CUHK)\\
$^3$Centre for Complexity Science and Department of Mathematics, Imperial College London\\
$^4$Helmholtz Institute for Functional Marine Biodiversity (HIFMB), Oldenburg\\
$^5$University of Oldenburg, Institute of Chemistry and Biology of the Marine Environment (ICBM), Oldenburg\\
$^6$Alfred-Wegener Institute, Helmholtz Center for Marine and Polar Research, Bremerhaven\\
$^7$Tourism College, Beijing Union University\\
$^\dag$These authors contributed equally.\\
$^*$gezhixiu@gmail.com
}

\date{\today}

\begin{abstract}
  Monitoring urban structure and development requires high-quality data at high spatiotemporal resolution. While traditional censuses have provided foundational insights into demographic and socioeconomic aspects of urban life, their pace may not always align with the pace of urban development. To complement these traditional methods, we explore the potential of analyzing alternative big-data sources, such as human mobility data. However, these often noisy and unstructured big data pose new challenges. Here we propose a method to extract meaningful explanatory variables and classifications from such data. Using movement data from Beijing, which are produced as a byproduct of mobile communication, we show that meaningful features can be extracted, revealing, for example, the emergence and absorption of subcentres. This method allows the analysis of urban dynamics at a high spatial resolution (here, 500m) and near real-time frequency, and high computational efficiency, which is especially suitable for tracing event-driven mobility changes and their impact on urban structures.
\end{abstract}

\maketitle

\section{Introduction}

Understanding the dynamics of cities is a central goal of urban studies. A variety of data-driven models have offered insights into the evolution of urban structures, focusing on diverse socioeconomic observables including income inequality~\cite{atkinson2009data, bryan2008evolution}, ethnic identities~\cite{kertzer2002censuses, benjamin2005evolution}, and environmental impacts~\cite{carpio2020influence,wang2024investigating}. For example, polycentric transitions are conceptualized as outcomes of competition between areas, defined by their economic allure and traffic congestion~\cite{louf2013modeling}. Similarly, urban scaling laws have been delineated through a balance between socioeconomic outputs and infrastructural costs~\cite{bettencourt2007growth}.

Traditionally, the development of cities has been studied using a variety of methods and data sources, including census datasets\cite{hammel1996model, sen2003pune, bromley2007new}. For instance, decadal censuses, such as the UK census, provide comprehensive information on an array of social variables such as education outcomes, employment status, and housing conditions, gathered from population surveys and aggregated spatially. Complementary to these are non-census datasets, including the American Communities Survey~\cite{USCensusBureau} and the UK's Indicators of Multiple Deprivations~\cite{payne2012uk}. Despite offering high-quality data from exhaustive surveys, the significant cost and time involved mean that census and similar datasets are released at long time intervals, thus offering only periodic snapshots of urban evolution. Additionally, relying on predetermined question catalogs makes these types of data less effective in identifying unanticipated developments.

To uncover emergent developments, analysis of real-time and alternative data sources is desirable. For instance, Germany has utilized open-source mobility data to analyze social structures and contact patterns during the COVID-19 pandemic~\cite{wiedermann2022evidence}. The introduction of high-frequency mobility data has enabled rapid analysis using unstructured and noisy, yet rich and comparatively unbiased, datasets, revealing the critical and diverse urban structures on much shorter timescales, e.g., the spatial and temporal decomposition of visitation~\cite{peng2012collective}, the impact of cultural ties on human mobility~\cite{wu2016geography}, and the nexus between contact patterns and epidemic propagation~\cite{barrat2021effect, oliver2020mobile}.

Mobility datasets are an incidental byproduct of our modern interconnected society. For example in mobile communications mobility traces are produced as a byproduct of the normal operations of network providers. Because the movement of individuals often occurs as a result of social needs, mobility data contains a wealth of information on social geography. However, as this data is not produced for this purpose, it only implicitly contains the social information. Careful data analysis is therefore required to extract salient social variables from mobility traces. 

In the analysis of tabular census-like datasets, recent progress has been made using diffusion maps~\cite{coifman2005geometric, coifman2006diffusion}, a manifold learning technique that reduces the dimensionality of structured datasets in biological and social studies~\cite{levin2021insights,ryabov2022estimation,fahimipour2020mapping,barter2019manifold}. Diffusion maps provide a nonlinear, deterministic, and hypothesis-free approach that pinpoints explanatory parameters in large high-dimensional datasets. For example, diffusion maps were recently used to extract explanatory variables from census data of specific cities and countries~\cite{barter2019manifold, xiu2023unravelling}, spotlighting higher education and deprivation hubs as key factors shaping their urban environment. The idea behind manifold-learning methods such as the diffusion map is that current datasets record much more information than is necessary to encode the salient information. The diffusion map can therefore reduce the number of variables by identifying the main variables that are needed to span the variation of data in the dataset. 

Here we propose the Mobility Census (MC), a computational framework for high-frequency analysis of urban structure. We start with a dataset of mobility traces that we segment into a 500m spatial grid. For each grid cell, we then compute a set of 1,665 different \emph{mobility variables} from the available traces. We work on the assumption that if a sufficiently large catalog of such variables is computed then the desired social information will become encoded in the resulting data table. We then use diffusion mapping to reduce the dimensionality again and extract a set of aggregated variables that account for the majority of the variance between cells and thus make the social information accessible in distinct variables.  

Using multi-year high-frequency mobility data of Beijing as an example, we discover the polycentric isolation patterns and separate local and global mobility features by analyzing indicators. Using additional data, we can interpret the eigenfeatures found by the diffusion map, and identify economic prosperity, location, and local irreplaceability as the most important mobility variables. Furthermore, we trace Beijing's accelerated evolution, including the evolution of subcentres from the functional supplements of the main city to independent entities. In some instances, this transformation can be attributed to substantial events like new airport construction, while in others, it is the cumulative effect of numerous smaller-scale changes. Thus, this study captures the dynamics of modern urban environments, paving the way for more nuanced understandings of city structures.

\section{A census for human mobility}

\begin{figure*}
  \centering
  \includegraphics[width=0.99\linewidth]{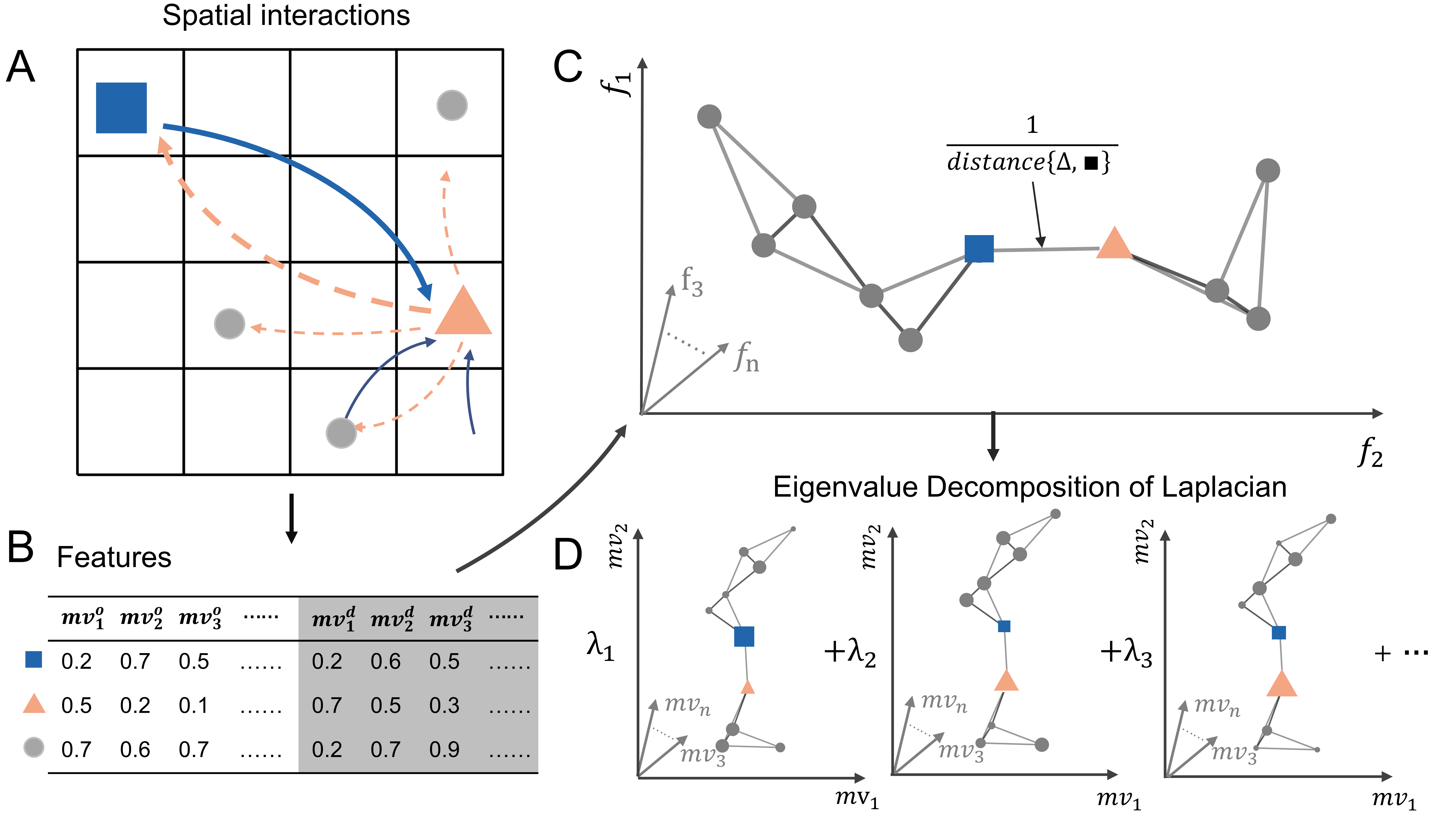}
  \caption{Sketch of the Mobility Census framework. The blue squares, gray circles, and orange triangles label the cells, while the arrows represent human mobility flows. The widths of these arrows denote the frequency of visits. \textbf{A}. Flow data are aggregated in a high-resolution grid with temporal and spatial granularity of 1 hour and 500m respectively. \textbf{B}. A large set (here 1,665) of statistical properties are calculated for each grid cell for each hour of the day, resulting in a high-dimensional data table. This table is then aggregated to provide a monthly summary of mobility patterns. \textbf{C}. To reduce the dimensionality the diffusion map is used, which is the first step constructing a network in which each spatial cell is connected to the $k$ most similar cells, and links are weighted by the respective distance. \textbf{D}. Finally, eigenvectors of a Laplacian matrix describing the data are computed. These eigenvectors assign new variables to the cells, providing a meaningful low-dimensional dataset parameterization. Subsequently, common analysis tools can be applied to this representation of the city.}
  \label{fig:sketch}
\end{figure*}

The mobility census method is a productional generalization of manifold learning by setting up a protocol first to aggregate the individual trajectories through each small area into a table of ``mobility variables'', then to apply diffusion mapping (DM) analysis to map the urban structures through DM eigenfeatures. The method is based on simple intuitions: a limited number of functional place categories influence human movements. Hence these categories should become encoded in movement traces, and thus can be extracted by a suitable analysis.  

To show the application of the Mobility Census method, we use a dataset containing movements of all China Unicom subscribers in Beijing from August 1-31, 2018, and May 1-31, 2021, amounting to ca.~\(11.57\times {10}^6\) users and \(1.8\times {10}^9\) trips, where a trip is an individual's single visitation from an origin to a destination. China Unicom is one of the three major ICT providers in China and Beijing, whose trip data has provided insights into many socioeconomic aspects such as tourism and local imbalanced developments~\cite{wu2020comparison, liu2021did, liu2023quantifying, louail2014mobile}. We note that already one month of data is sufficient to reveal key elements of the evolving urban structure (see below). Moreover, we verified that the coverage rate of the China Unicom does not have a significant spatial bias in terms of districts (see Supplementary Information, \textit{SI} Figure S1), and thus should provide a reasonably unbiased view of the spatial structure of the city. 

We partition the area of Beijing by a grid of \(500{\rm m}\times 500{\rm m}\) cells (number of cells $N = 22,704$). For each cell, movements originating or concluding within it are identified, resulting in variable-length lists of timestamped movements (Fig.\ref{fig:sketch}A). These movements are then represented in the form of an Origin-Destination matrix for each respective hour, organized by both origin and destination cells. Acknowledging the potential influence of the Modifiable Areal Unit Problem (MAUP) on our results, we performed a sensitivity analysis by re-partitioning the area into a 1km$\times$1km grid, and compared the results derived from 500m and 1km grid. This sensitivity analysis showed that our primary observations were consistent across different grid sizes. However, the impact of more localized, detailed activities and the significance of specific, less obvious patterns (e.g. neighborhood-wise home-work segregation that is distinct within a 1km scale) varied with the change in grid size. The larger grid analysis mostly confirmed our initial findings at \(500{\rm m}\times 500{\rm m}\) cells, particularly for broad spatial patterns like commuting and night-time activities. However, it also highlighted finer distinctions in small-scale patterns such as the delineation of residential and work areas. Despite these differences, our main findings based on the 500m grid remained robust, illustrating the general properties of human mobility and nuanced patterns at a community scale (500m).

To reduce the complexity of the dataset, we aim to identify the essential \textit{features} and their combinations that shape human movements. We collect characteristic statistical attributes indicative of an area's movement, hereinafter referred to as \textit{mobility variables} (Fig.~\ref{fig:sketch}B). These mobility variables cover a full range of topics from existing literature that associates the movement properties with urban developments, e.g., the number of trips originating from the area, the total distance of all trips, or the average speed of movement within the area on weekdays. Each of the mobility variables quantifies the collective traits of the trips that start or terminate in the respective cell (precise definitions of the mobility variables in \textit{SI} Table S1). Furthermore, we incorporate certain geostatistical operators (e.g., the H-index and Gini coefficient, see \textit{SI}) to the basic statistics to cope with the human mobility partially driven by the places' comprehensive functions~\cite{pappalardo2015returners}. The additional operators help to reveal the nonlinear responses of location attractiveness to human movements. In this manner, we generate a census-like feature table, offering a fixed dimensionality of 1,665 mobility variables for each cell. This breadth of mobility variables prevents overreliance on a limited set of variables, which is particularly important in complex urban settings.

Constructing the feature table (Fig.~\ref{fig:sketch}B) brings structure and a first reduction of data complexity, but the feature table is still a high-dimensional dataset that suffers from the curse of dimensionality~\cite{weber1998quantitative}. To recover the most dominant factors determining the attractiveness of locations, we then explore this table using a diffusion map analysis (Fig.~\ref{fig:sketch}C and D) that was previously applied to census data~\cite{barter2019manifold}. The basic idea of the diffusion map~\cite{coifman2005geometric} is that salient features of the data can be discovered by analyzing the topological structure of the dataset. A central insight underlying the diffusion map is that comparisons between very dissimilar objects are highly unreliable and introduce noise that can quickly swamp the salient information. It is therefore essential to remove such low-confidence comparisons of cells from the analysis. The analysis starts by finding the most similar pairs of cells. Following \cite{ryabov2022estimation}, we compute the similarities between cells as a Spearman rank correlation~\cite{spearman1961proof} between the cell's feature list (see \textit{SI}). Utilizing a proven approach~\cite{altman1992introduction, nadler2008finite}, we limit the comparisons used in the subsequent steps to the 10 most similar cells of each cell. 

The remaining comparisons of mobility features between cells now form a complex network~(Fig.~\ref{fig:sketch}C), that can be mathematically described by a row-normalized Laplacian matrix~\cite{barter2019manifold}. The dimension of the eigenvectors of this matrix equals the number of cells (Fig.~\ref{fig:sketch}D). Hence, each eigenvector of the Laplacian assigns a value to each of the cells. We can thus interpret the entries of each of the eigenvectors as a new feature for the cells. The features identified in this way are in many ways similar to principal components~\cite{abdi2010principal}, but provide a more robust, nonlinear parameterization of complex high-dimensional data. 

In the following, we refer to the new features identified from the DM as \textit{eigenfeatures} (EF). Each EF corresponds to an eigenvalue that scales inversely with the variation captured by the respective feature. Hence the eigenvalues are indicative of the importance of the respective features, such that the most important eigenfeature is the one with the lowest non-zero eigenvalue. We note that the DF analysis identifies important statistical patterns but does not provide an interpretation of these patterns. 
Instead, we use two approaches to help us formulate hypothesis regarding these patterns: First we can visualize important eigenfeatures on a map by color-coding grid cells according to the value of the respective eigenfeature (Fig.~\ref{fig:E_distribution}). Second, we correlate the eigenfeatures with the original mobility variables to identify the mobility variables to which a particular eigenfeature is linked.

\section{Dominant Patterns}
\begin{figure*}
  \centering
  \includegraphics[width=0.99\linewidth]{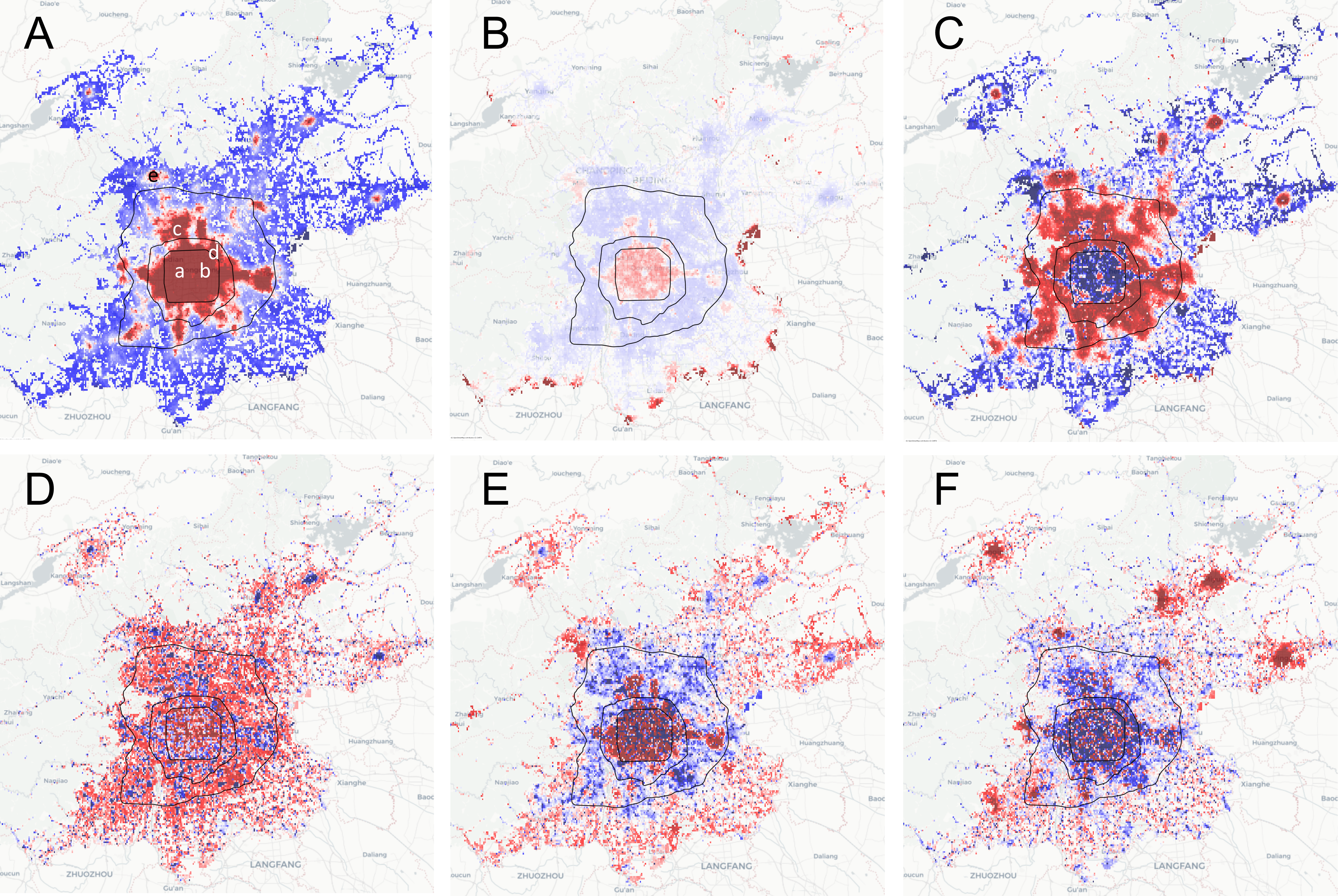}
  \caption{Explanatory variables (eigenfeatures, EF) identified by the mobility census are depicted in Panels A-F, presented in descending order of importance. Values are represented by a gradient where shades of red correspond to more positive entries and shades of blue denote more negative entries. We interpret the EFs operationally as indicators of centrality (A), entry points to the city (B), local heterogeneity (C), livability (D) workplaceness (E), and attractiveness for long-distance journeys (F). Labeled places in A are  Financial Street (a), China World Trade Centre (b), Xierqi-Huilongguan subdistrict (c), Wangjing subdistrict (d), and Changping town (e). These demonstrate that diffusion mapping can identify informative features in the data.}
  \label{fig:E_distribution}
\end{figure*}

We start our analysis by plotting the most important EFs of Beijing that were derived from the mobility data of the year 2021, (Fig.~\ref{fig:E_distribution}) color-coded by the EF's entries, and calculate the correlation of the eigenfeatures with original mobility variables (most correlated ones in Tab~\ref{tab:corrtab}). We claim that most interpretations of the eigenfeatures are consistent between the mobility census of the year 2018 and 2021 with few exceptions, primarily due to pronounced event-driven changes in visitations (discussed further in Section~\ref{sec:subcentre}). This provides some indication of the robustness of the mobility census results.

\begin{table*}
  \centering
  \small
  \begin{tabular}{cccccccccccccc}
  \toprule
  \textbf{Rank} & \textbf{\( f_1 \)} & \textbf{Corr} & \textbf{\( f_2 \)} & \textbf{Corr} & \textbf{\( f_3 \)} & \textbf{Corr} \\
  1 & H-index (commute) & 0.8843 & Gini (Flow ratio (16)) & 0.7238 & Gini(H-index) & 0.5636 \\
  2 & out-flow (15) & 0.8621 & Gini (Flow ratio (12)) & 0.6957 & Gini (ROG(w,p4)) & 0.5330 \\
  3 & out-flow (12) & 0.8613 & Gini (Flow ratio (17)) & 0.6810 & Gini (ROG(w,p3)) & 0.5300 \\
  4 & out-flow (16) & 0.8610 & Gini (stay duration (h,p8)) & 0.6763 & Gini (ROG(w,p5)) & 0.5245 \\
  5 & out-flow (total) & 0.8605 & Gini (Flow ratio (13)) & 0.6691 & Gini (travel Dis(w,p6)) & 0.5132 \\
  \textbf{Rank} & \textbf{\( f_4 \)} & \textbf{Corr} & \textbf{\( f_5 \)} & \textbf{Corr} & \textbf{\( f_6 \)} & \textbf{Corr} \\
  1 & LR (in-flow(10)) & 0.8419 & net commute flow & 0.4989 & Average travelling time (w,p4) & 0.2892 \\
  2 & LR (r-population) & 0.8265 & in-flow (8) & 0.4943 & Average travelling time (w,p5) & 0.2881 \\
  3 & LR (in-flow(2)) & 0.8190 & in-flow (9) & 0.4873 & Average travelling time (w,p3) & 0.2839 \\
  4 & LR (in-flow(21)) & 0.8177 & entropy of work & 0.4588 & Average travelling time (w,p6) & 0.2797 \\
  5 & LR (in-flow(20)) & 0.8175 & LR (in-flow(8)) & 0.4451 & Average travelling time (w,p2) & 0.2726 \\
  \end{tabular}
  \caption{Correlations between mobility variables and diffusion map eigenvectors. Gini(\( \cdot \)) represents the Gini coefficient applied to a variable within its 2km neighborhood. LR(\( \cdot \)) denotes the ratio of a cell's variable value to the mean value of that variable across the cell's 2km neighborhood. K is the kurtosis of the distribution. The notation (w, p1) is the first percentile of a characteristic of a cell as destination, while \( \cdot \)(h, p2) is the second percentile of a cell's characteristics as origin. \( (t) \) specifies the mobility metric at \( t \) hours since midnight.}
  \label{tab:corrtab}
\end{table*}

For the first EF, $f_1$,  we find the highest values in the centre, which coincide with central business areas such as Financial Street (a), China World Trade Centre Towers (b), and Sanlitun. Pronounced local maxima also occur at emerging hubs of economic activity such as Xierqi-Huilongguan Fig.~\ref{fig:E_distribution}A(c) and Wangjing, Fig.~\ref{fig:E_distribution}A(d), well-known as the headquarters of most high-paying, high-tech companies, which act as local hubs of development. We thus conclude that $f_1$ detects a high density of workplaces in the urban centre and subcentres.

To further explore $f_1$ we find the most strongly correlated mobility variables. The strongest correlations with indicators of a high volume of flow toward areas ranked highly by $f_1$ ($0.86$, $P<0.001$). This is consistent with our interpretation as the 1\% of cells that score highest in this indicator contain 6\% of the residential population but 13\% of the workplaces. Also highly correlated is an indicator of flow diversity~($0.57$, $P<0.001$), which indicates that the areas highlighted by $f_1$ receive flow from a diverse range of origins. 

The second EF, $f_2$, is the most strongly localized of the first six EFs, which can be mathematically verified by computing the inverse participation ratio (see \textit{SI}). It has pronounced maxima at several locations in the south where major highways, such as the G103, G106, G230, and G102, enter the city. Moreover, we find a maximum in the centre of Beijing at Sanlitun, an area well known for its embassies and nightclubs. What unites these locations is that they receive significant long-distance travel at nighttime hours, which is due to late-night party-goers (Sanlitun), or trucks, which are not allowed to travel in the daytime under Chinese regulations (motorway entry points). The long-distance, nighttime visits create a distinct traffic pattern that the DM picks up. The most correlated variable with $f_2$ is an indicator of the diversity of in- and out-flow ($0.72$, $P<0.001$) and trip duration ($0.68$, $P<0.001$), which is consistent with this interpretation. 

In 2018, $f_2$ also highlighted some areas in the northern subcentres (see \textit{SI}), but the respective maxima are no longer visible on the map for 2021.  It can be interpreted as a sign that the subcentres have lost attractiveness as long-distance destinations in this period.  Indeed, the house-job ratio and residential population in these subcentres increased significantly~\cite{lin2015impact} and hence likely receive less long-range commuter traffic. 

EF $f_3$ has a pronounced concentric structure, with strong positive values found both in the city centre and outlying villages, whereas the outer areas of the city are assigned negative values. In diffusion maps, such high-low-high patterns can appear as harmonic modes of other prominent features. One must therefore particularly careful to avoid over-interpreting them. However, in a real date, even harmonic modes often convey useful information. 

Considering the metrics that correlate with $f_3$ highlight an indicative measure which we refer to as the `diversity of centrality values,' mathematically represented by the Gini coefficient, calculated on the h-index, where the h-index is defined as the count of destinations in a target cell's neighborhood that each have a flow volume exceeding $h$. This indicator underscores the variation in the importance, or centrality, of the neighboring destinations around a particular location. Thus places receiving high values in this EF are those surrounded by locations that differ in importance. Such differences are very pronounced in the city centre, whereas the outer areas supporting the centre are much more uniform. In the outermost belt strong differences return, likely due to the spatial self-organization of outlying villages \cite{christaller1966central}. Hence, the boundary line where $f_3$ crosses from the negative back into the positive can be regarded as the true boundary of the city.

\begin{figure*}
  \centering
  \includegraphics[width=0.99\linewidth]{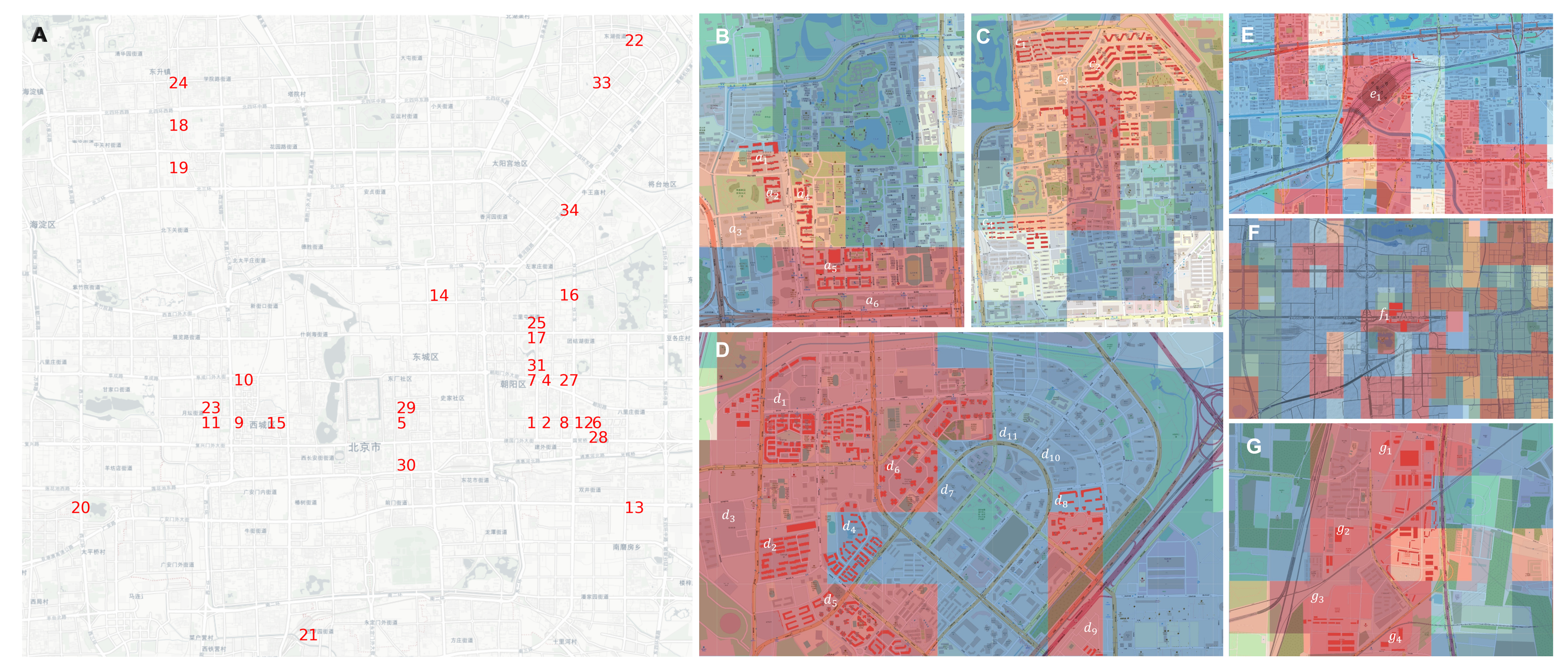}
  \caption{Small-scale patterns of the EFs determined by the spatial transitions of EF's extreme values. \textbf{A.} Blue cells correspond to the largest 100 entries of the differences $f_3-f_1$, in the central part of Beijing (Fifth Ring Road). Red numbers are top-searched shopping malls retrieved from Google Map API. \textbf{B-D.} Peking University, Tsinghua University, and Wangjing. Cells are colored by the entries of $f_5$ as in Fig.~2E, thus red for positive entries and blue for negative entries. Low transparent red and blue highlight the uses of buildings, such as dormitory/residential (red), and teaching/office buildings (blue). \textbf{E-G.} South, and West Railway Station, and Xinfadi wholesale food market, colored by the entries of $f_6$ from the most negative (blue) to the most positive (red). High values of $f_6$ correspond to areas that are visited by visitors from distant origins. Specific labels of locations are listed in \textit{SI} Table S3. These results illustrate that the mobility census reveals some insights down to the 500m scale. }
  \label{fig:local}
\end{figure*}

Computing the difference between the $f_3$ (indicating variance of centrality) and $f_1$ (indicating centrality) highlights local places of interest. To confirm this we compared the highest values of $f_3-f_1$ in the centre to the most searched shopping malls which reveals a very good agreement (Fig.~\ref{fig:local}A). 

The next three EFs have a pronounced structure on the 500m scale. 
EF $f_4$ correlates very well with sinks and sources of short-range commuter traffic, with positive (negative) values marking the sources (sinks) of flows (Fig.~\ref{fig:E_distribution}D). The EF also correlates strongly with mobility variables that measure the relative volume during hours that correspond to typical closing times of businesses, corroborating this interpretation (e.g., with correlation coefficient $0.82$, $P<0.01$ with in-flow at 8 p.m.). We see that emerging software industry centres at Xierqi, Wangjing, and Yizhuang all receive strongly positive values. EF $f_5$ is similar but correlates with morning opening hours rather than evening closing times (e.g., with correlation coefficient $0.49$, $P<0.01$ with in-flow at 8 a.m.). These observations suggest $f_4$ and $f_5$ being livability and workplaceness indicators, respectively.

To corroborate the interpretation of $f_5$, we also explored it on a smaller scale by considering the locations of Peking University, Tsinghua University, and Wangjing. These locations are identified by the largest average differences of the $f_5$'s entries with their neighbors. The detailed scale $f_5$ separates workplaces and residential areas within these areas (\ref{fig:local}B-D). 

EF $f_6$ also exhibits a highly detailed pattern with positive and negative values occurring often in close proximity. However, the centre receives mostly positive values, whereas the subcentres have mostly negative entries. This eigenfeature correlates strongly with mobility variables indicating long-distance trips. Hence we interpret this eigenfeature as an indicator of long-distance attractivity. It is confirmed by considering the entries on the detailed scale where the highest values of this EF are found at railway stations and the largest wholesale food market (Fig.~\ref{fig:local}E-G). 

Interestingly, repeating the analysis for 2018 (Fig.~S10) also reveals pronounced positive values in the sub-centres, which have vanished by 2021. It could indicate a change in mobility behavior induced by the COVID-19 restrictions, which also constrained travel on this scale and/or the increasing residential population mentioned above.  

\section{Subcentre evolution} 
\label{sec:subcentre}
\begin{figure*}[t]
  \centering
  \includegraphics[width=\linewidth]{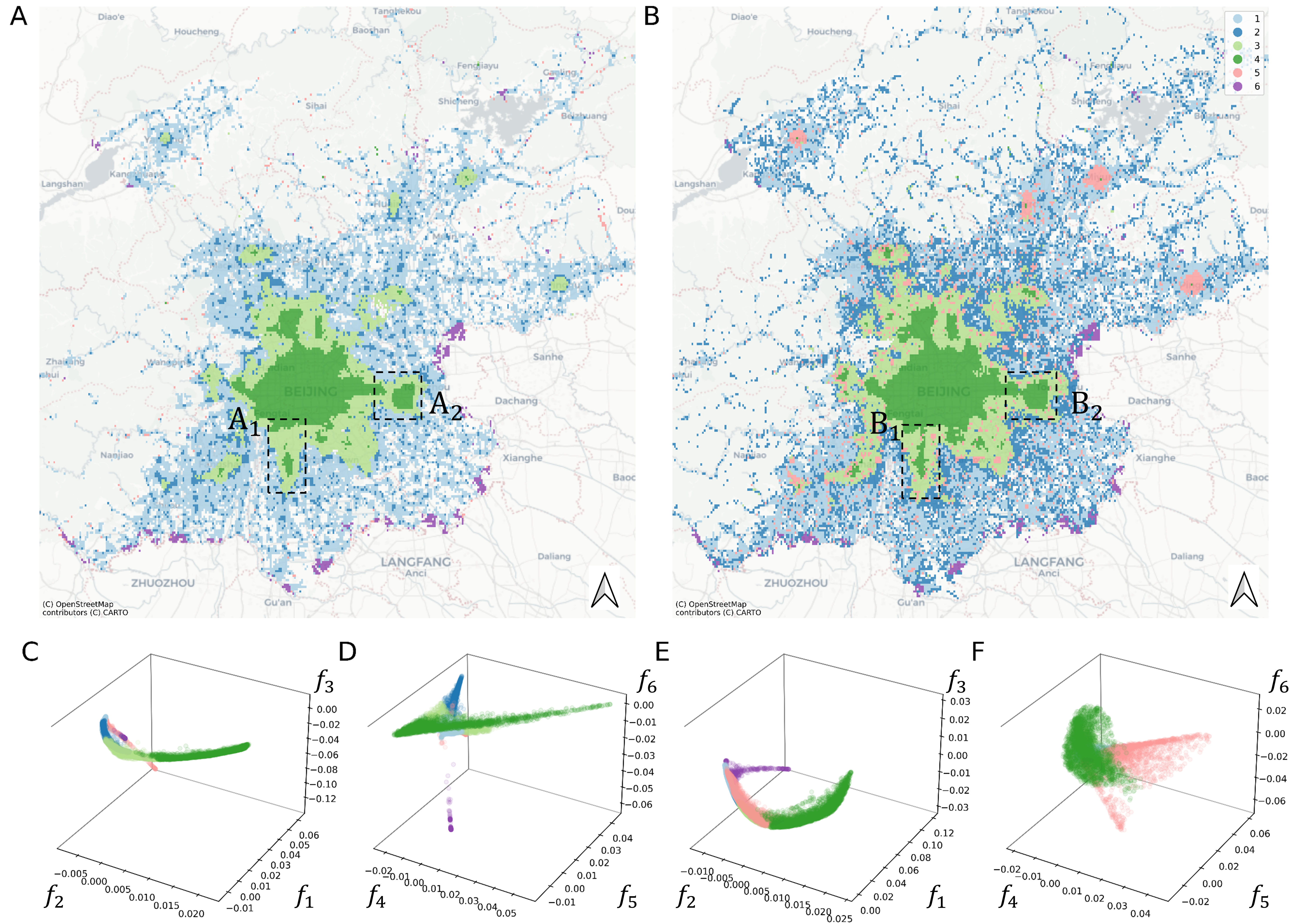}
  \caption{Classification of areas and urban development.
  Clustering results are displayed both on maps (A, B) and in a data space spanned by three of the eigenfeatures, EF 1, 2, and 3 (C for 2018 and E for 2021) and EF 4, 5, and 6 (D for 2018 and F for 2021). Results are shown for one month in 2018 (A, E, F) and one month in 2021 (B, C, D). Labels make Daxing (1) and Tongzhou (2) in the maps. The results show that nicely coherent clusters are obtained (colors, 1-6), which identify distinct functional areas of the city. The longitudinal comparison illustrates the emergence of distinct subcentres in the north and the absorption of subcentres at Daxing and Tongzhou.}
  \label{fig:cluster}
\end{figure*}
Above we showed that the DM can extract salient functional variables (the EFs) from the high-dimensional set of mobility variables. It thus provides a reduction of the dimensionality of the data that is also valuable for subsequent analysis. The EFs effectively reduce the dimensionality of our data and highlight key mobility patterns within the city. However, each EF represents a specific aspect of urban mobility, and considering them individually might not provide a complete or coherent picture of the overall structure of the city. Additionally, the sheer number of EFs can make it challenging to identify overarching patterns or to compare different regions of the city. 

Here we further aggregate the data by applying a Gaussian mixture model (GMM)~\cite{levin2021insights}, a statistical model, that can be used to break the data into distinct clusters. In the Beijing data, GMM identifies 6 clusters (see \textit{SI}) representing 6 types of areas distinct by similar mobility properties. 

To gain a visual impression of the quality of the clustering result, we can visualize the clusters in the data space defined by the most important EFs 
(Fig.~\ref{fig:cluster}C-F). This visualization shows the partition of the data manifold into coherent sections. 
Coloring the clusters in geographical space (Fig.~\ref{fig:cluster}A,B) reveals a clear separation into different areas, which we can operationally identify as rural areas (clusters 1,2, depending on local centrality), urban fringe (3), urban centre (4), subcentres (5), and major gateways to the city (6).

We now use these operational designations in a longitudinal comparison of the situation in 2018 and 2021. While the big picture in both of these years is similar, there are some notable differences. First, we notice that the category that we identified as a subcentre is much more prominent in 2021 than it was in 2018.  
During this period, three areas in Northeastern Beijing (Pinggu, Huairou, and Miyun) and one area in Northwestern Beijing (Yanqing) transitioned from the urban fringe to the subcentre category. 
We conclude that new work opportunities, a rising residential population, and also possibly COVID-19-related mobility restrictions have caused these four areas to develop into fully-fledged subcentres, which was also confirmed by field research from literature~\cite{dai2022land, li2020evolution}.

Another area of interest is Daxing in southern Beijing ($A_1$ in Fig.~\ref{fig:cluster}A and $B_1$ in Fig.~\ref{fig:cluster}B). In 2018, this well-developed subcentre is classified as an urban centre while being separated from the main city centre by an area of the urban fringe. By contrast in 2021, an area of 39 grids ($\sim 9$ km$^2$) that covers the central area of Daxing has become connected to the main city centre. A major event in this area that occurred in the intervening period is the opening of Beijing Daxing International Airport. We conclude that the construction and opening of this airport tied Daxing closer to the city centre, which is also evidenced by the construction of major motorways and underground connections in this area. As a result, the subcentre of Daxing was effectively absorbed into the city centre. 

We see a similar development also in Tongzhou~(43 differently classified cells from $A_2$ in Fig.~\ref{fig:cluster}A to $B_2$ in Fig.~\ref{fig:cluster}B, $\sim  $ 10.75 km$^2$), which becomes likewise connected to the city centre between 2018 and 2021. In this case, the development was likely triggered by the relocation of the Beijing municipal government to Tongzhou in 2019.

\section{Conclusions}\label{sec12}
In this paper, we proposed a new method, the mobility census, for the analysis of urban structure from big unstructured datasets. The proposed method first generates a large number of different metrics (here 1,665 mobility variables, elaborated in \textit{SI} Section II) for each geographical area, to turn the unstructured dataset into a structured table. We then use the diffusion map to extract a smaller number of salient features. This reduces the dimensionality of the data, and thus avoids the ``curse of dimensionality'' while enabling subsequent analysis. Beyond the particular application considered here, other unstructured data sources could be analyzed using the same approach: breaking the domain of interest into small units, compiling a large table of statistical features for these units, and using diffusion mapping to extract comprehensive features.

The primary limitation of the mobility census is its focus on active individuals, neglecting the city's vulnerable groups. This oversight can lead to a skewed understanding of urban dynamics, as it fails to capture the mobility challenges of less mobile or less connected populations such as the elderly, disabled, or economically disadvantaged. 

By contrast, the major advantages of the mobility census are that it can reuse data that is already available, reducing costs and workload. It provides results very fast on a near-real-time basis, requiring few weeks of data and negligible processing time, which opens up the option to keep pace with urban development while it happens. Finally it avoids reliance on a narrow question catalog, which enables the discovery of novel features not anticipated by the researcher.  

Application of the mobility census to Beijing showed that the method can identify distinct functional classes of areas. While the diffusion map does not in itself provide an interpretation of these classes, interpretations can be assigned using expert knowledge. We note that such interpretations, including those in this paper, should at first be treated as hypotheses, but can later be corroborated using additional analysis and data.   

Here this analysis identifies major explanatory variables that shape the city (cf.~\cite{barter2019manifold}), such as attractivity, workplace/housing density, and nighttime activity. 
Revealing these features provides insights into the functional organization of cities and their temporal evolution. Notably, the method provides this information with high spatial (here, 500m) and temporal (hourly basis, aggregated to 1 month collection of mobility variables) resolution. 

The dimensionality reduction provided by the diffusion map also enables subsequent steps, such as the clustering analysis presented here. We showed that this analysis provides a useful tool to categorize areas within cities and identify boundaries. Moreover, it provides a high-resolution view of important geographical processes, such as the emergence of fully-fledged subcentres and the absorption of subcentres into the city centre.  

The mobility data used in this study are presently produced at a massive scale as a byproduct of mobile communication. The mobility census method can be applied to aggregated data products of such mobility data, thus avoiding data protection concerns. Moreover, it provides a numerically efficient, deterministic, and hypothesis-free approach to the analysis. We envision that in the future, the application of this method may provide a high-resolution and near-real-time view, of the evolution of our ever-growing and ever-accelerating urban environments.

\section{Code Availability}

The code to derive mobility variables and diffusion maps is available in \href{https://github.com/GXIU/Mobility-Census}{https://github.com/GXIU/Mobility-Census}. The source data of anonymous users' mobile checking-in is accessible through a purchased license in China Unicom's server. We used SQL queries to aggregate the individual traces to the locations' 1,665 mobility variables which are accessible on GitHub when published. 

\bibliographystyle{unsrt}
\bibliography{ref}

\end{document}